\documentstyle[pre,aps,twocolumn,epsf,floats]{revtex}

\begin{document}

\draft

\twocolumn[\hsize\textwidth\columnwidth\hsize\csname@twocolumnfalse\endcsname

\title{Depinning of kinks in a Josephson-junction ratchet array}

\author{E. Tr\'{\i}as$^{1}$, J. J. Mazo$^{1,2}$, F. Falo$^{2}$ and
T. P. Orlando$^{1}$}

\address{$^{1}$ Department of Electrical Engineering and Computer
Science,\\ Massachusetts Institute of Technology, Cambridge,
Massachusetts 02139,\\} \address{$^{2}$ Departamento de F\'{\i}sica de
la Materia Condensada and ICMA \\ CSIC-Universidad de Zaragoza,
E-50009 Zaragoza, Spain}

\date{Submitted to PRE July 28, 1999}
\maketitle
\tightenlines

\begin{abstract}
We have measured the depinning of trapped kinks in a ratchet potential
using a fabricated circular array of Josephson junctions.
Our ratchet system consists of a
parallel array of junctions with alternating cell inductances and
junctions areas.  We have compared this ratchet array with other
circular arrays.
We find experimentally and numerically that the
depinning current depends on the direction of the applied current
in our ratchet ring.
We also find other properties of the depinning
current versus applied field, such
as a long period and a lack of reflection symmetry,
which we can explain analytically.
\end{abstract}


]

\section{Introduction}

Disorder and noise are not always undesirable in physical systems.
Inhomogeneity has been shown to control certain types of
spatiotemporal chaos\cite{braiman95}, while noise can lead to an
enhancement of the signal-to-noise ratio because of stochastic
resonance \cite{sr}.  Another more recent counterintuitive result
is that of transport of a Brownian particle in a ratchet potential
\cite{astumian97}.  Though initially proposed as a model for molecular
motors in biological organisms \cite{magnasco93}, ratchets
can also serve as a model to study dissipative and stochastic
processes in nanoscale devices.

\begin{figure}[t]
\epsfxsize=2.0in
\centering{\mbox{\epsfbox{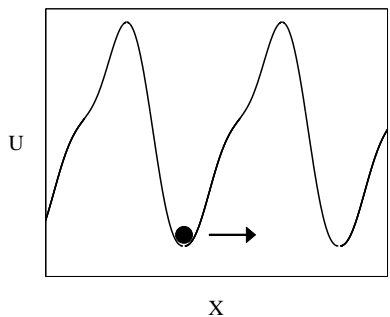}}}
\vspace{0.1in}
\caption[]{Example of a ratchet potential.  The particle sitting
on the well requires less force to move through the first
peak to the right than to move to the left.  Therefore, there
is a preferred direction of motion.
}
\label{fig:ratpot}
\end{figure}

A ratchet potential is a periodic potential
which lacks reflection symmetry (in 1D $V(x) \neq V(-x)$, see
Figure~\ref{fig:ratpot}).  A consequence of this symmetry
breaking is the possibility of rectifying non-thermal,
or time correlated, fluctuations
\cite{hangi}. 
This can be understood intuitively.
In Fig.~\ref{fig:ratpot}, it takes a smaller dc driving force 
to move a particle
from a well to the right than to the left.
In other words, the spatial symmetry of the dc force 
is broken. Under an ac drive
(so-called ``rocking ratchets'') or
time-correlated noise, particles show net directional
motion in the smallest slope direction.  This effect can be used in
devices in which selection of particle motion is desired. 

Because of this effect, ratchet engines have been proposed as devices 
for phase separation \cite{prost94},
and very recently as a method of flux cleaning in
superconducting thin films \cite{lee99}.  A ratchet mechanism
has also been proposed as a method to prevent mound formation in
epitaxial film growth \cite{derenyi98}.

Josephson junctions are solid state realizations of
a simple pendulum.  By coupling them,
it is possible to make a physical realization of model
systems such as the
 Frenkel-Kontorova model for dislocations \cite{FK,par}
or the 2-D  X-Y model \cite{resnick81} for
phase transitions.  In particular,
a  parallel Josephson array (see Fig.~\ref{fig:diagram})
is a discrete version of the sine-Gordon equation and it
has been used to experimentally study soliton (usually
referred to as kinks, vortices or fluxons) dynamics
on a discrete lattice \cite{par}.

In parallel arrays, kinks behave as particles in which 
the idea of Brownian rectification can apply.
The applied current is the driving force.
If the kink experiences a ratchet potential,
then the current needed to move the kink in
one direction is different than the current to move
it in the opposite direction.

In this paper, we will show that we can design 
almost any type of 1D pinning potential in
a parallel Josephson array by choosing an appropriate
combination of
junction critical currents and plaquette areas.
Indeed, it has been shown \cite{falo99} that two alternating
critical currents and plaquettes areas are enough to provide a ratchet
potential for fluxons.
As we will show below, this is not the only possible  design for 
a ratchet potential.

With only an ac driving current these arrays show dc voltage steps of stability 
at multiples of the external ac drive amplitude.  
This occurs when the equivalent ac driving force
becomes commensurate with the period of the ratchet potential.    
This behavior
could open the possibility of using these arrays for a voltage standard
device or a microwave detector without a dc bias current. Moreover, the
same ideas of flux cleaning underlying reference 
\cite{lee99} could be applied to 
2D arrays using the designs described here.

The paper is organized into 5 sections. 
Section \ref{sec:model} introduces the theoretical
framework for the study of an inhomogeneous parallel Josephson arrays.  We
find that inhomogeneous arrays present a long periodicity
with respect to the number of kinks in the array.  To
test the theory, we have designed four different Josephson junction rings and
measured the depinning current of the array versus the applied 
magnetic field. The experimental results are shown in section \ref{sec:exp}.
In section \ref{sec:disc} we discuss some of the properties of the model and
show that they agree well the experimental results.  We also show that a combination of
three different critical current junctions is sufficient to design a
ratchet potential.  In section \ref{sec:conc} we present the conclusions of our work
and propose a number of new experiments.

\section{Theoretical framework}
\label{sec:model}

\subsection{Circuit model}

\begin{figure}[b]
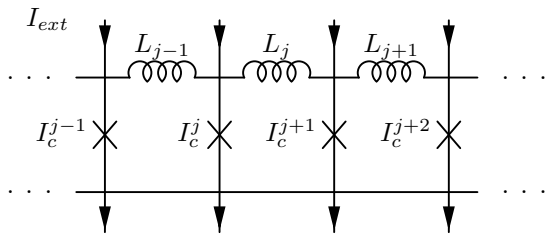

\begin{center}
\input fig2.tex
\ \box\graph
\end{center}
\caption{Circuit diagram for an inhomogeneous parallel Josephson array.  Each junction
has a critical current $I_c^j$ and each cell has an inductance of
$L_j$.}
\label{fig:diagram}
\end{figure}

Figure~\ref{fig:diagram} shows the circuit diagram for an array of
Josephson junctions.  Each junction is marked by an ``$\times$'' and
we will connect $N$ junctions in parallel with short wires as shown.
Coupling of the junctions occurs through the geometrical inductances
of the cells. We will neglect all mutual inductances and consider only
the self-inductance of each cell $L_j$.  The induced flux in each cell
is then $L_j$ times the mesh current of the cell which in this simple
geometry can be easily seen to equal the current through the top horizontal
link $I^j_b$.  We will use $I_{ext}$ for the uniformly
applied external bias current per junction as shown in 
Fig.~\ref{fig:diagram}.
We then define the mesh current as the current passing
through this top horizontal wire.  With this definition we can place
the loop self-inductance $L_j$ on the top horizontal link.  We
emphasize that this inductance is not the wire inductance, but the
self-inductance of the cell so that only one such element is needed
per cell.

The junctions will be modeled by the parallel combination of an ideal
Josephson junction with a critical current of $I_c^j$, a capacitor
$C_j$, and a resistance $R_j$.  The ideal Josephson junction has a
constitutive relation of $I_j=I_c^j\sin \varphi_j$ where $\varphi_j$
is the gauge-invariant phase difference of the junction.  When 
there is a voltage across the junction, $v_j$, then
$v_j=(\Phi_0 / 2 \pi) d\varphi_j / dt$.  Since we
will have $N$ parallel junctions, in our array $j=1$ to $N$.

The circuit equations result from applying current conservation
and flux quantization\cite{fq}.
Current conservation at the top node of junction $j$ yields
\begin{equation}
C_j\dot{v}_j+{v_j \over R_j}+I_c^j\sin\varphi_j=I_{ext}+I_b^j-I_b^{j-1}
\end{equation}
Flux quantization of cell $j$ yields
\begin{equation}
{\Phi_0 \over 2\pi}(\varphi_{j+1}-\varphi_j)=\Phi_j,
\label{eq:fq}
\end{equation}
where $\Phi_j$ is the total flux in cell $j$.

Due to the linearity of Maxwell's equations, $\Phi_j$ can be
decomposed into two parts: the induced flux $\Phi_{ind}^j$, and the
external flux $\Phi_{ext}^j$ which is the applied field $B_{ext}$ times the
cell area $A_j$.  The induced flux is simply $L_j$ times the mesh
current of the cell, which has been defined to equal $I_b^j$.  Then,
\begin{eqnarray}
C_j\dot{v}_j & + & {v_j \over R_j}+I_c^j\sin\varphi_j 
= I_{ext}  + F_j \nonumber \\
& & +{\Phi_0 \over 2 \pi} \left[{1\over L_j}(\varphi_{j+1}-\varphi_j) + 
{1\over L_{j-1}}(\varphi_{j-1}-\varphi_{j})\right]
\end{eqnarray}
with $F_j=(\Phi_{ext}^{j-1}/L_{j-1} - \Phi_{ext}^{j}/L_{j})$.  

This circuit is realizable by varying cell and junction areas.
The cell area $A_j$ will determine the self-inductance.  If $W$ is
the width of the cell and $\Delta x_j$ is its length
then $L_j \approx \mu_0 \Delta x_j$ as long as 
$W\sim \Delta x_j$.
Since $\Phi_{ext}^j=W\Delta x_j B_{ext}$, we see that 
$\Phi_{ext}^{j}/L_{j}\approx W B_{ext}/\mu_0$ and is
approximately constant for all $j$.
The junction area determines $I_c^j,
C_j$, and $R_j$ but they are not independent since the capacitance and
critical current are linearly proportional to the junction area and
the resistance is inversely proportional to the junction area. 
The $I_c^j R_j$ product and the $I_c^j/C_j$ ratio of each junction are
constant for every junction.

We will normalize all the currents by $I_c^\star=\max(I_c^j)$ and
time by $\tau=\sqrt{ \Phi_0 C_\star / 2\pi I_c^\star}$ 
where $C_\star=\max(C_j)$. 
Then,
\begin{eqnarray}
h_j {\cal N}(\varphi_j)= & &i_{ext} + f_j  \nonumber \\
& &+ \lambda_j(\varphi_{j+1}-\varphi_j)+
\lambda_{j-1}(\varphi_{j-1}-\varphi_{j})
\label{eq:insge}
\end{eqnarray}
where
${\cal N}(\varphi_j)=\ddot{\varphi}_j+\Gamma\dot{\varphi}_j+\sin\varphi_j$
\cite{w1}.
The ratio of critical currents is $h_j=I_c^j/I_c^\star$ and the inductances
are normalized as $\lambda_j=\Phi_0/2\pi I_c^\star L_j$.
Finally, $f_j=2\pi f(\lambda_{j-1} A_{j-1}/A_{\star}-\lambda_j A_j/A_{\star})$, 
where $f$ is the frustration $B_{ext}A_{\star}/\Phi_0$.
We have used $A_\star=\max(A_j)$.

To complete the system we need to specify the boundary
conditions. There are two types: open, if the junctions form
a linear row; and periodic, if the junctions form a closed ring. 

For the open boundary condition we set $\lambda_{0}=A_0=0$
in Eq.~\ref{eq:insge} for junction $j=1$. 
At the other end of the array, $j=N$, we set 
$\lambda_{N}=A_{N}=0$.

For the periodic
boundary conditions we let $\lambda_{0}=\lambda_{N}$ and $A_0=A_N$.
Furthermore, a circular 
system poses a topological constraint on $\varphi_j$
since they are angular variables and have $2\pi$
periodicity: $\varphi_{j+N}=\varphi_j+2\pi M$.
In particular $\varphi_0=\varphi_{N}-2\pi M$ and 
$\varphi_{N+1}=\varphi_{1}+2\pi M$.  Here $M$
is referred to as the winding number and represents the number
of kinks in the system.

In this paper we will discuss systems with periodic boundary
conditions. Since the product $\lambda_j A_j$ is roughly constant
throughout the array we consider $f_j=0$ in the simulations of
the rings we present \cite{open}. We have checked numerically that
for the experiments reported here, these
terms do not significantly alter our results.

\subsection{Symmetries}
\label{sec:symm}

The system of equations (\ref{eq:insge}) presents an odd inversion symmetry
under the change $M \to -M$,  $\varphi_i \to -\varphi_i,$ and 
$i_{ext} \to -i_{ext}$ as is expected from Maxwell's equations. 
The response of the array to an external
current will reflect this symmetry.  In particular, 
$I_{dep}(-M)=-I_{dep}(M)$.  Here,
$I_{dep}$ is 
the maximum value of the
applied current for which a solution $\dot{\varphi_j}=0$ can not be sustained
in the presence of a positive or negative external current.
To refer to this odd inversion symmetry we
will use the notation 
$I_{dep-}(-M)=I_{dep+}(M)$
where $I_{dep \pm}$ refers to the {\it absolute value} of the 
depinning current as the external current is increased or decreased
from zero.

Another symmetry of the equations refers to the periodicity of the
system when varying the number of kinks in the array. In the case of a
regular ring (all the cells and junctions are equal) this period $T$
is equal to the number of junctions, $N$ \cite{sg}. 

A method of calculating the periodicity in $M$ for the general case 
studied here is to use the simple
transformation
\begin{eqnarray}
\psi_j=\varphi_j + 2\pi m_j
\label{eq:transf}
\end{eqnarray}
where $m_j$ are integers. The equations of motion in the new
variables are the following
\begin{eqnarray}
h_j{\cal N}(\psi_j -  2\pi m_j) =
\lambda_j(\psi_{j+1}-\psi_j)+  
\lambda_{j-1}(\psi_{j-1}-\psi_{j})  \nonumber  \\
- 2 \pi \lambda_j(m_{j+1}-m_j)
- 2 \pi \lambda_{j-1}(m_{j-1}-m_{j}) + i_{ext} + f_j
\label{eq:transeq}
\end{eqnarray}
where ${\cal N}(\psi_j -  2\pi m_j)={\cal N}(\psi_j)$. The
new boundary conditions are
\begin{equation}
\psi_{j+N}=\psi_j+2\pi (M + T)
\end{equation}
where $T=m_{j+N}-m_j$. 
Thus after the transformation (\ref{eq:transf}) we recover the same
equations as (\ref{eq:insge}) but with the number of kinks equal to $M +
T$ so that the equations are periodic in the number of kinks in
the array with a period $T$.

To calculate $T$
we take out the $m_j$ dependence on the
right hand side of Eq.~\ref{eq:transeq} by choosing
$m_j$ such that 
$\lambda_j(m_{j+1}-m_j)+\lambda_{j-1}(m_{j-1}-m_{j})=0$. 
Remarkably, the resulting period is independent of $h_j$ and
only depends on the
ratio between the consecutive $\lambda's$. 
In the appendix we find a 
formula for the periodicity in the number of kinks for
the general system.

Here we are going to develop the case of a ring that was measured:
a ring with
an even number of junctions and with alternating cell areas.
In this case there are only two $\lambda's$ involved. Let
$\lambda_j=\lambda_1 (\lambda_2)$ for $j$ odd(even) 
and $\lambda_1/\lambda_2=p/q$.  If we let
$(m_{j-1}-m_j)=-q$ and $(m_{j+1}-m_{j})=p$ (for even $j$
for instance), we satisfy the above condition.

The period is calculated from the new boundary conditions:
\begin{equation}
T=m_{N+1}-m_1= (p+q)N/2.
\label{eq:per}
\end{equation}
For the regular
array $p=q=1$ and we recover the expected result of $T=N$. 
Also, we note that in
order to have a finite period
we need the ratios between $\lambda's$ to be rational numbers. 
This condition will almost never be satisfied in
a real experiment. Thus we see that a simple design of
alternating cell areas can result in an arbitrarily long
period (that could be equal
to $\infty$) when varying the number of kinks in the array.

A similar calculation can be made for the case of an open array. As no
topological constraint for the phases can be imposed, the number of
kinks in the array does not appear in our equations. We consider instead
the periodicity of the system with the external field.  In
this case, the periodicity depends on the ratio between the cell areas
instead of the ratio between the inductances.  It can be shown that
the period in $f=B_{ext}A_\star/\Phi_0$ is equal to $q$, where
$A_2/A_1=p/q$ and $A_1=A_\star$.

\section{Experimental results}
\label{sec:exp}

\begin{figure}[t]
\epsfxsize=2in
\centering{\mbox{\epsfbox{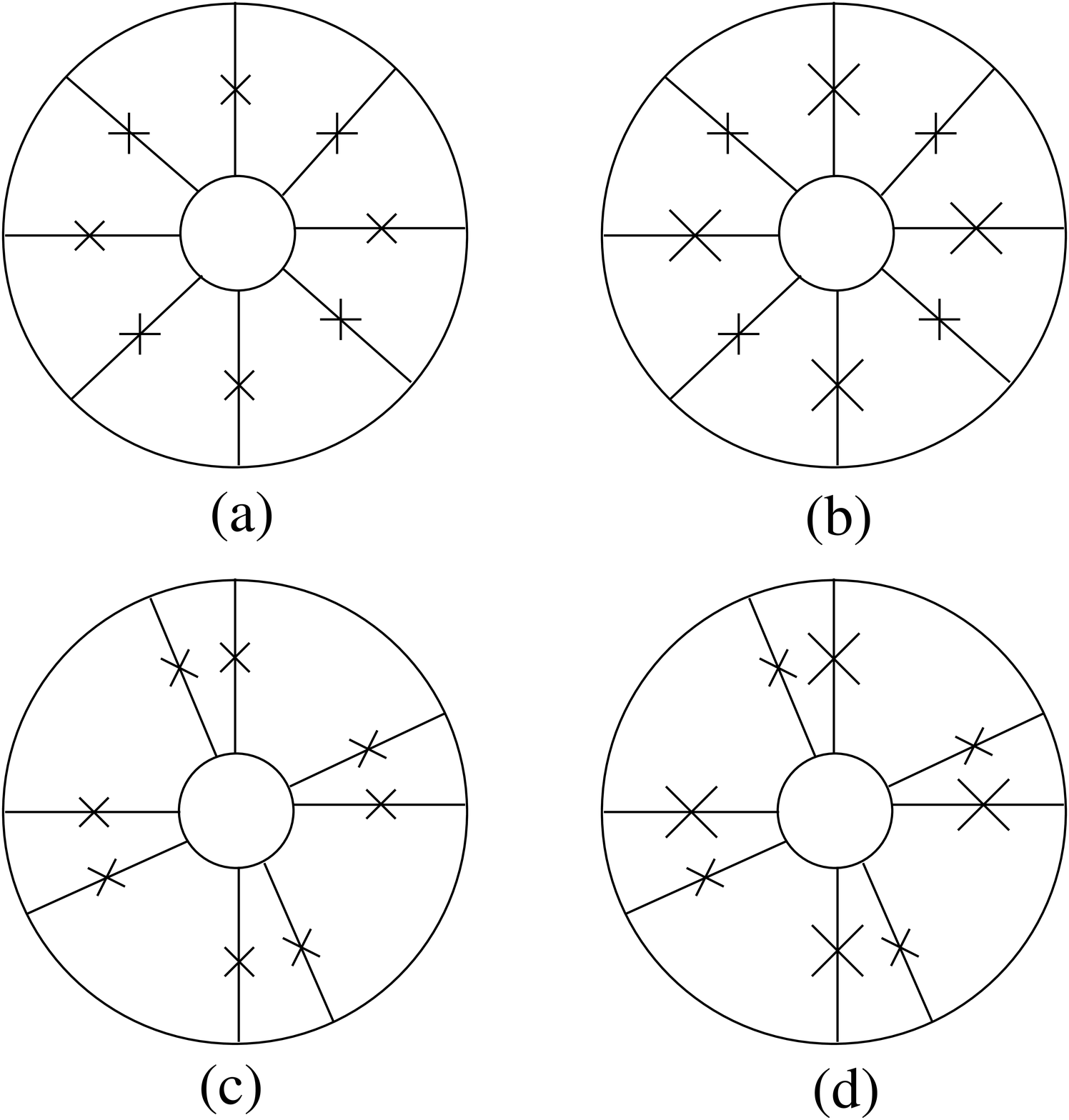}}}
\vspace{0.1in}
\caption[]{The four different measured arrays: (a) regular ring, 
$h_j=1$ and $\lambda_j=0.11$
(b) ring with alternating critical currents, $h_j=1$ and $h_{j+1}=0.43$,
$\lambda_j=0.043$,
(c) ring with with alternating cell area, $h_j=1$, $\lambda_j=0.08$
and $\lambda_{j+1}=0.15$
(d) ratchet ring with alternating critical currents and cell areas,
$h_j=1$, $h_{j+1}=0.43$, $\lambda_j=0.035$,
and $\lambda_{j+1}=0.06$.  These parameters are calculated
at $T=0\,K$.
}
\label{fig:all}
\end{figure}

We have designed and fabricated the four different rings (a), (b), (c)
and (d) schematically shown in Fig.~\ref{fig:all}. The rings are
fabricated with a Nb-Al$_2$O$_x$-Nb tri-layer technology with a
junction critical current density of $1 \, {\rm kA/cm^2}$.  The
current is injected through bias resistors in order to
be distributed as uniformly as possible.  We measure the dc voltage
across a single junction \cite{dcv} and each ring consists of $N=8$
junctions. 

Fig.~\ref{fig:all}(a) is a regular ring with equal critical 
currents and plaquette areas.  Fig.~\ref{fig:all}(b) 
has alternating critical currents with a ratio of 0.43.
Fig.~\ref{fig:all}(c) has alternating plaquette areas with a 
ratio of $\lambda's$ of 1.8.  Finally, Fig.~\ref{fig:all}(d)
has both alternating critical currents and 
alternating plaquette areas.  It will be shown 
experimentally that only (d) has a ratchet pinning potential.

The outer diameter of each ring is $36\,{\rm \mu m}$ with
an area $\sim 4070\,{\rm \mu m}^2$.  The
inner diameter is $18\,{\rm \mu m}$ and it consists of
an island of niobium that is used to extract the applied
current.
The rings also have either
small junctions ($3 \times 3\,{\rm \mu m^2}$) or alternating
small and large junctions ($4.25 \times 4.25\,{\rm \mu m^2}$).
The designed $I_c$ ratio is 0.5, but in practice 
the junction areas have rounded corners and experimentally
we find the $I_c$ ratio to be 0.43.
We vary the cell
inductance by alternating the cell area.  In this case,
the angles of the cells
are $60^{\circ}$ and $30^{\circ}$.  

Both $\Gamma$ and $\lambda$ are mostly determined from 
material properties of the samples and the junction $I_c$.
Since $I_c$ varies with temperature, both parameters can be
experimentally controlled to some extent.  In general $\Gamma$ and
$\lambda$ can be made larger by up to a factor of 10 by
raising the sample temperature.  As the
temperature reaches $T_c$, however, most of the 
measured features become too smeared to be distinguished.

The temperature dependence of $I_c$ is modeled well by the standard
Ambegaokar-Baratoff relation with $I_{c}(0)R_{n} = 1.9\,{\rm mV}$
\cite{ambegaokar63}.  We find that $I_c(0)=95\,{\rm \mu A}$ for the
small junctions and $I_c(0)=224\,{\rm \mu A}$ for the larger
junctions.  We will normalize all our parameters with the largest
$I_c$ of a given ring.  From the above values, we can estimate $\rm
\Gamma(0)=0.17$ which, due to the constant $I_cR_n$ product, is
independent of junction area.  The inductances are estimated from a
numerical package that extracts inductances from complex
3-D geometries of conductors \cite{fasthenry}.
In this sample the loop
inductance is ${L} = 23.5\,{\rm pH}$ for the small cells and ${L} =
42.6\,{\rm pH}$ for the large ones [arrays (c) and (d)].  For the
cells in rings (a) and (b) ${L} = 33.5\,{\rm pH}$.  To
calculate dimensionless penetration depth $\lambda(0)=\Phi_{0}/2\pi L
I_c(0)$ we use $I_c=95\,{\rm \mu A}$ if the ring only has
small junctions [(a) and (c)] and for those rings that also have large
junction [(b) and (d)] we use  $224\,{\rm \mu A}$.

The current-voltage, IV,
curves are measured by applying a perpendicular magnetic field
of 0 to 300 mG through a magnetic coil that is mounted on the
radiation shield of our probe.  We heat the sample above
$T_c=9.2\,{\rm K}$ and cool down to a temperature $T<T_c$. We cool our
ring in the presence of a flux that corresponds to approximately $M$
flux quanta.  Flux quantization will cause the expulsion of extra flux
so that the ring contains exactly $M$ flux quanta after undergoing a
superconducting transition.

\begin{figure}[tb]
\epsfxsize=3in
\centering{\mbox{\epsfbox{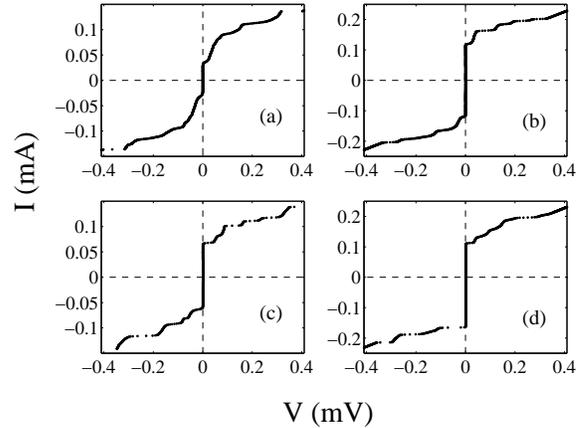}}}
\vspace{0.1in}
\caption[]{Sample IV curves for the four rings considered
in {\protect Fig.~\ref{fig:all}}  ((a) 
corresponds to {\protect Fig.~\ref{fig:all}}(a) and
so on).  Rings (a), (b), and (c)
have symmetric IV's as the current is swept in the positive
and negative direction.  The measurements correspond
to $M=1$.   Ring (d) is the ratchet ring as
can be seen from the difference in the depinning current
in the positive and negative direction.
}
\label{fig:iv}
\end{figure}

Figure~\ref{fig:iv} shows typical IV's for the
different rings shown in Fig.~\ref{fig:all}.  Fig.~(a)
is for a regular ring when $M=1$.  The IV is symmetric with 
respect to applied current direction.  As the
current is increased from the superconducting state
the voltage remains at zero.  We define the depinning current
when the array has a voltage greater than a threshold
of $1.5\,\mu {\rm V}$.  Our computer controlled equipment
also corrects for any voltage drift of our amplifiers.
As the current increases
beyond the depinning value, there
is a sequence of voltage steps where as the current 
increases the voltage remains relatively constant.
There are at least two mechanisms that can cause
these steps: resonances between 
the circulating kink and radiated linear waves,
and instabilities of the whirling branch\cite{par}.  We have 
verified that the voltage positions correspond
to these two mechanisms.  

Figure~\ref{fig:iv}(b) is a ring
with alternating critical currents when $M=1$.  We again
see that the IV is symmetric with respect to
current direction and that there
are voltage steps.  These steps are of the
same origin as in the regular ring.  However,
in this ring the linear dispersion relation that
determines the resonance condition is split
into two branches.  This splitting is analogous
to the optical and acoustic branches of a crystal
with a two atom basis.   Fig.~(c) is a ring
with alternating areas.  The characteristics
are similar to that of ring (b) including a splitting
of the linear dispersion relation.  Since for these
three rings $I_{dep+}=I_{dep-}$, we can infer that the kink
is traveling in a symmetric pinning potential as
theoretically expected.

Figure \ref{fig:iv}(d) shows an IV for
the ring with both alternating critical currents
and areas.  The IV of this ring is qualitatively
different from the other rings due to the ratchet
nature of the pinning potential.  We see that $I_{dep}$
in the positive direction is 
$\sim 65\%$ of the depinning current in the negative direction.  We
also note that there are different voltage
steps excited in the up and down direction.  
The steps are of the same nature as the explained resonances
above and there is also a splitting of the dispersion relation.
In the rest of the article we will focus on $I_{dep}$
measurements as a signature for ratchet behavior
in our arrays.

\begin{figure}[tb]
\epsfxsize=3in
\centering{\mbox{\epsfbox{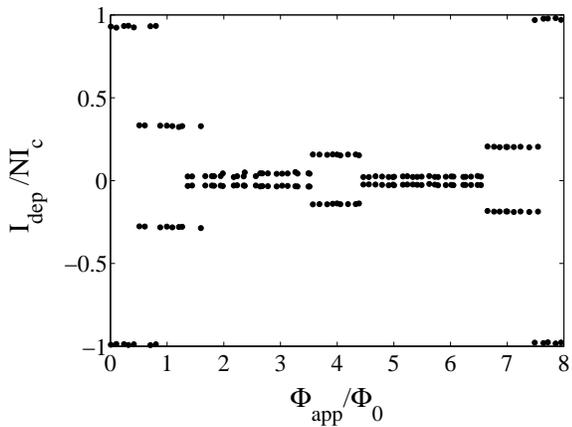}}}
\vspace{0.1in}
\caption[]{Measured critical currents vs. applied flux
for a regular ring. To
calculate applied flux, we multiply the applied field by 
the ring area.  A constant offset has also been subtracted to account
for the ambient magnetic field. The measurement was
done at $T=8.8\,{\rm K}$ with $\Gamma \approx 0.5$ and $\lambda \approx 0.9$.
}
\label{fig:ring_icf}
\end{figure}

Figure~\ref{fig:ring_icf} shows a measurement of the depinning current
vs. applied flux for the regular ring shown in
Fig.~\ref{fig:all}(a).  The temperature is $8.8\,{\rm K}$,
$\Gamma=0.5$ while $\lambda =0.9$.  Each plateau represents a
different number of kinks trapped in the ring.  This is a direct
result of flux quantization: The ring only allows integer number of
flux quanta even if we have applied slightly more or less flux.  Since
$N=8$ and this ring has a symmetric pinning potential, we expect 
$I_{dep+}= I_{dep-}$ (no ratchet effect), and a
period of 8 as can be
seen in the measurements.  We also see that $I_{dep}$ has a
reflection symmetry about $M=T/2$.

\begin{figure}[tb]
\epsfxsize=3in
\centering{\mbox{\epsfbox{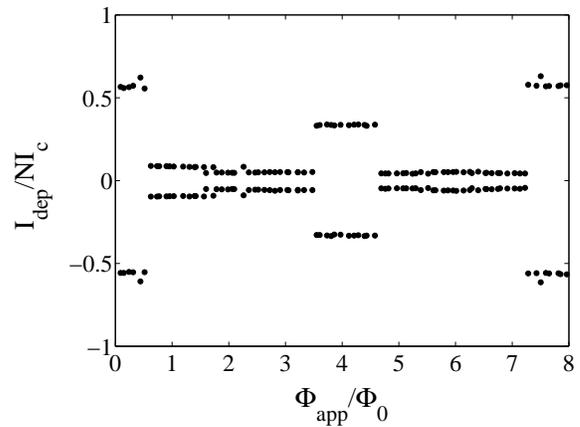}}}
\vspace{0.1in}
\caption[]{Measured critical currents vs. applied flux
for a ring with alternating critical currents. The
applied flux was calculated as described in {\protect Fig.~\ref{fig:ring_icf}}.
The measurement was
done at $T=9\,{\rm K}$ with $\Gamma \approx 0.7$ and $\lambda \approx 0.9$.
}
\label{fig:ring_icf_ic}
\end{figure}

When we alternate the critical currents in our ring we expect 
the same qualitative features of $I_{dep}$ as in the regular ring.
Figure~\ref{fig:ring_icf_ic} shows a measurement of the depinning current
vs. applied flux for the ring shown  in Fig.~\ref{fig:all}(b)
which has alternating critical currents.  There are plateaus corresponding 
to different values of $M$ just as in the regular ring and
there is up-down symmetry and periodicity with $M=8$ as expected,
and a reflection symmetry about $M=4$.

\begin{figure}[tb]
\epsfxsize=3in
\centering{\mbox{\epsfbox{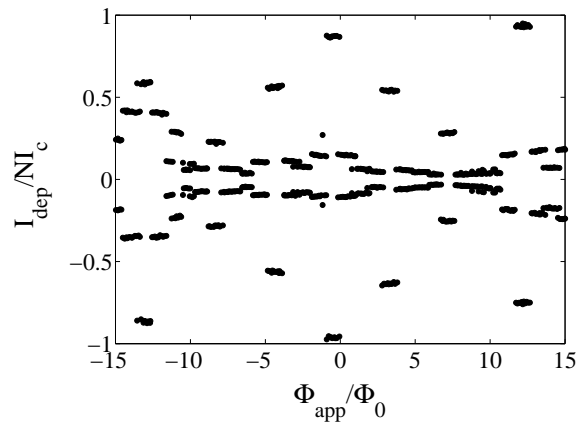}}}
\vspace{0.1in}
\caption[]{Measured critical currents vs. applied flux
for a ring with alternating cell areas. The
applied flux was calculated as described in {\protect Fig.~\ref{fig:ring_icf}}.
The measurement was
done at $T=9\,{\rm K}$ with $\Gamma \approx 0.7$ and $\lambda_l \approx 0.7$
and $\lambda_s \approx 1.3$.
}
\label{fig:ring_icf_area}
\end{figure}

If we make all the critical currents constant and vary only the cell area
as in Fig.~\ref{fig:all}(c), then we alternate the values of $\lambda$ but
the pinning potential remains symmetric.
At $T=9\,{\rm K}$, $\lambda_l$ for the large cell is $\approx 0.7$ and
$\lambda_s$ for the small cell is $\approx 1.3$.  The result of
measuring $I_{dep}$ is shown in Fig.~\ref{fig:ring_icf_area}.  
As expected the data is symmetric with respect to current direction so
kinks are not traveling in a ratchet pinning potential.  However, unlike in the
previous rings, $I_{dep}$ is no longer periodic with $M=8$.  As shown
in section \ref{sec:symm}, the period will depend on the ratio of the
inductances.  For our geometry $L_1/L_2\approx 1.8$ or $9/5$ which
implies a period of 56.  However, in any physical array the inductance
ratio is rarely going to be exactly a ratio of small numbers.  Just
on physical grounds we expect a very large period, if any, in the
experiments.  In Fig.~\ref{fig:ring_icf_area} we have measured the
depinning current from $M=-15$ to $M=15$ and though there is some
apparent self-similarity in the data, it is not periodic.  Though
there is no period, we can still prepare our ring systematically with
$M=1$, 2, 3, etc. by counting the plateaus.  But instead of $M=1$ and
$M=1+N$ yielding the same dynamical system as in the regular ring,
they are now distinguishable.

\begin{figure}[tb]
\epsfxsize=3in
\centering{\mbox{\epsfbox{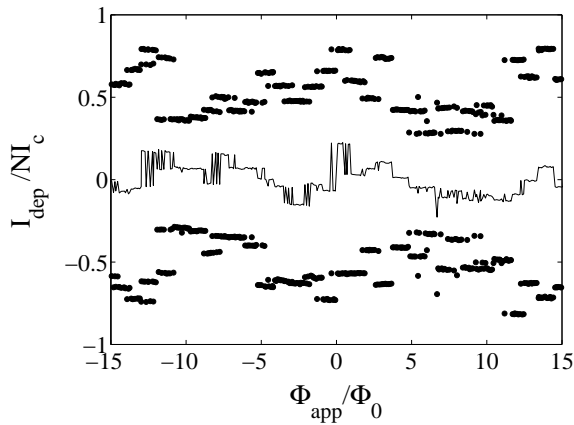}}}
\vspace{0.1in}
\caption[]{Measured critical currents vs. applied flux
for a ratchet ring. The
applied flux was calculated as described in {\protect Fig.~\ref{fig:ring_icf}}.
The measurement was
done at $T=8.8\,{\rm K}$ with $\Gamma \approx 0.5$ and $\lambda_l \approx 0.3$
and $\lambda_s \approx 0.6$.  The line varying about
$I_{dep}=0$ is the difference between $I_{dep+}$ and
$I_{dep-}$.
}
\label{fig:ring_icf_rat}
\end{figure}

When we alternate both the critical current and the cell inductances
as in Fig.~\ref{fig:all}(d), it is possible to form a ratchet
pinning potential (see Figure~\ref{fig:ratpot}).
Fig.~\ref{fig:ring_icf_rat} shows an experiment on such a ring.  Since
the period depends on the inductance ratio, we experimentally expect a
very long period.  This is borne out by the data as there is no sign
of a period in the range from $M=-15$ to 15. 
We also expect that $I_{dep+}\neq I_{dep-}$
since the kink is traveling in a ratchet pinning potential.
The line shown in the center of the figure varying about
$I_{dep}=0$ is the difference between the $I_{dep+}$ and $I_{dep-}$.
Clearly, the force to move kinks in one direction is different than
the force to move it in the opposite direction.  The magnitude and
direction of this ratchet effect depends on the number of kinks in the
system.

As a further test of the symmetries and periods of
the experiments, we have numerically integrated Eq.~\ref{eq:insge}
using a variable step size explicit
4th order Runge-Kutta method.
The kink number $M$ is set in
the boundary junctions.  The initial conditions are
$\varphi_j=2\pi Mj/ N$. That is, we stretch the kinks across the full
array at the start of the simulation.  We then sweep the applied
current in the positive direction until a voltage develops
in the array and calculate $I_{dep+}$.  We repeat the
procedure while sweeping the current in the negative direction
to calculate $I_{dep-}$.

\begin{figure}[tb]
\epsfxsize=3in
\centering{\mbox{\epsfbox{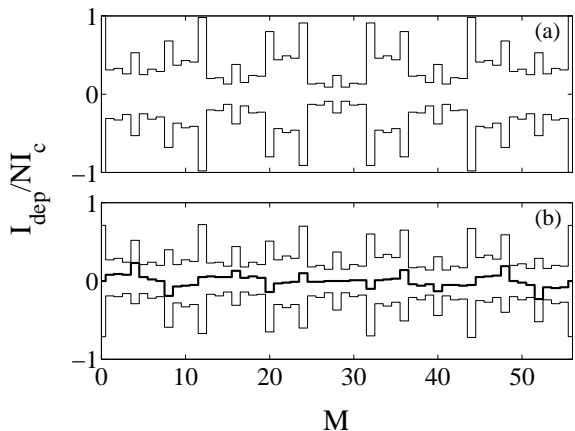}}}
\vspace{0.1in}
\caption[]{(a) Simulation for $N=8$ ring with
only alternating $\lambda$'s of $\lambda_l=0.3$ and $\lambda_s=0.54$,
and $h_j=1$.
(b) Simulation for $N=8$ ratchet ring with both alternating $\lambda$'s
of $\lambda_{j}=0.3$ and $\lambda_{j+1}=0.54$ and critical currents of
$h_j=1$ and $h_{j+1}=0.43$.
Line shown about $I_{dep}=0$ is the difference
between $I_{dep+}$ and $I_{dep-}$.
}
\label{fig:sim_dep}
\end{figure}

Figure~\ref{fig:sim_dep} shows the simulations with
parameters similar to those of the experiments.
Both Fig.~\ref{fig:sim_dep}(a) and
Fig.~\ref{fig:sim_dep}(b) have alternating $\lambda's$
with $\lambda_j=0.3$ and $\lambda_{j+1}=0.54$ for $j$ odd. The inductance
ratio is $0.54/0.3=9/5$ so using Eq.~\ref{eq:per} the 
expected period is $T=56$.  
We find this period in the simulations.
Fig.~\ref{fig:sim_dep}(a) has $h_j=1$ so we expect the
depinning current to be up-down symmetric, i.e. no ratchet effect,
as can be seen in the data.
Since we always have an odd inversion symmetry 
[$I_{dep+}(M)=I_{dep-}(-M)=I_{dep-}(T-M)$],
$I_{dep}$ is symmetric about $M=56/2=28$.  This 
reflection symmetry of $I_{dep}$ about $T/2$ is generic for 
any array that is not ratchet since it is a direct 
consequence of the up-down symmetry of the currents.  We
also find this symmetry in the experiments of
non-ratchet arrays.

Fig.~\ref{fig:sim_dep}(b) has junctions with two alternating
critical currents ($h_j=1$ and $h_{j+1}=0.43$ for $j$ odd)
as well as two alternating $\lambda$'s
($\lambda_j=0.3$ and $\lambda_{j+1}=0.54$ for $j$ odd).
We now expect
the kinks to travel in a ratchet pinning potential 
so that $I_{dep+}$ does not equal $I_{dep-}$, though $I_{dep}$ still
has an odd inversion symmetry.  Just as in the experiments
we see that the effect of the ratchet, and
rectification direction,  depends
on the number of kinks.  Also, $I_{dep}$
does not have the expected reflection symmetry about 
$T/2$.  In summary,
the simulations show the 
same features as the experiments and also agree quantitatively
with our predictions.

\section{Discussion}
\label{sec:disc}

The equations developed in section \ref{sec:model}
describe kink propagation
through a discrete inhomogeneous medium.   In this
section we will try to get a better understanding
of the system by briefly analyzing the continuous
limit of our discrete equations.  We will then
go back to our discrete equations and approximate
the pinning potential for a single kink.  With
the analysis, it will become 
apparent how it is possible to construct many types of pinning 
potentials, including ratchet ones, in the inhomogeneous array.

To derive the
continuous limit of the equations, let
$2{\bar\lambda}_j=\lambda_{j}+\lambda_{j-1}$ and
$\delta\lambda_j=\lambda_j-\lambda_{j-1}$.  Substituting in
Eq.~\ref{eq:insge}, we get
\begin{equation}
h_j{\cal N}(\varphi_j)={\bar\lambda}_j\partial_{xx}\varphi_j
+\delta\lambda_j\partial_x\varphi_j + f_j + i_{ext}
\end{equation}
where $\partial_{xx}\varphi_j=\varphi_{j+1}-2\varphi_j+\varphi_{j-1}$
represents a discrete Laplacian while
$\partial_x\varphi_j=(\varphi_{j+1}-\varphi_{j-1})/2$ is
just the center difference of the first order derivative.  
To arrive at a continuous limit we expand our variables as 
Taylor series in $\Delta x_j$.  The cell area is $W\Delta x_j$
while the cell inductance $L_j=G\Delta x_j$ 
as $\Delta x_j \to 0$ where
$G$ is a geometric constant.  Therefore 
$f_j=0$ as $\Delta x_j \to 0$
and the discrete operators are replaced by 
their continuous  derivatives
\begin{eqnarray}
h(x) {\cal N}(\varphi) &= &  \lambda(x)\partial_{xx}\varphi + \partial_{x}\lambda(x)\partial_x \varphi +i_{ext} \nonumber \\
 &= &  {\partial_x } \left( \lambda(x){\partial_x }\varphi \right) +i_{ext}  
\end{eqnarray}
If $\lambda$ and $h$ are constant then we have the usual sine-Gordon
equation.  In this case the equations have a reflection symmetry and
it is not possible to have a ratchet pinning potential. If $\lambda$ is
dependent on position, the spatial coupling is analogous to
inhomogeneous diffusion, anisotropic heat conduction, or waves
traveling in an anisotropic medium.  We also note that $f_j$ in the
discrete equations is essentially a perturbation to the continuous
model that is dependent on the exact discretization employed and is
usually small.  Thus, in order to get a ratchet pinning potential, there are
three ways to break the reflection symmetry of the equations: with an
appropriate $h(x)$, $\lambda(x)$, or a combination of both.

To calculate how the parameters $h_j$ and $\lambda_j$
determine the pinning potential,
we will use a perturbative approach.   In the limit
where all $\lambda_j \to 0$ the kink will approach a step
function \cite{baesens}.  
A stable kink configuration will have the kink
sitting in a potential well in the middle of a plaquette.
Let the kink lie between junction
$j$ and $j+1$.  The nearest phases to $j$ and $j+1$ will
be small in this limit.
As an approximation we let
$\varphi_j=\alpha$ and
$\varphi_{j+1}=2\pi-\beta$
and set all the  other phases  to 0
or $2\pi$.  
We can solve for $\alpha$ and $\beta$ by
minimizing the static energy of the system,
\begin{equation}
H=\sum_j \left[{\lambda_j\over 2}(\varphi_{j+1}-\varphi_j)^2+h_j(1-\cos\varphi_j
)\right].
\label{eq:dsgen}
\end{equation}
Here we 
have ignored the kinetic energy since 
we are only concerned with kink depinning \cite{jung96}.

Substituting we are left with
\begin{eqnarray}
H &= & {1 \over 2}({\lambda_j }+{\lambda_{j+1} })\beta^2
+{1 \over 2}({\lambda_{j-1}}+{\lambda_j })\alpha^2 \nonumber \\
& & -2\pi\lambda_j(\alpha +\beta)
+\lambda_j\alpha\beta
+2\pi^2\lambda_j \nonumber \\
& & +h_j(1-\cos\alpha)+ h_{j+1}(1-\cos\beta).
\end{eqnarray}

To solve for $\alpha$ and $\beta$ we minimize the energy:
$\partial H / \partial \alpha=0$ and
$\partial H / \partial \beta=0$.  The resulting
equation is transcendental because it depends on the
sine of $\alpha$ and $\beta$ and
would in general have to be solved numerically.  However,
for the systems of small $\lambda's$ studied here,
the corrections are small and we can
linearize
the sine terms ($\sin(x)\approx x$) to solve for $\alpha$ and $\beta$. 
We have found that for the parameters used in this paper the
linear approximation is sufficiently accurate to describe the numerically
calculated pinning potentials.

After linearizing the sine term we are left with
\begin{eqnarray}
\alpha&=&{2\lambda_j\pi }(h_{j+1}+\lambda_{j+1})/D \nonumber \\
\beta&=&{2 \lambda_j\pi } (h_j+\lambda_{j-1})/D,
\end{eqnarray}
where
$D=(h_j+\lambda_j+\lambda_{j-1})(h_{j+1}+\lambda_{j+1}+\lambda_{j})-\lambda_j^2$.

To get an idea of how the energy depends on the
parameters, we
can substitute back into Eq.~\ref{eq:dsgen} and
expand the energy as a series with respect to $\lambda_j$.
The result is
\begin{equation}
H  =  2\pi^2\lambda_j +O(\lambda_j^2)
\end{equation}

For small $\lambda_j$, the height of the pinning potential
when the kink is the middle of a plaquette
is determined by $\lambda_j$.  The second order term
has corrections due to $h_{j-1}$, $h_{j}$ and $\lambda_{j-1}$
and $\lambda_{j+1}$.

As the kink moves through the pinning potential
it will reach a point of maximum energy which
in the limit where all
$\lambda_j \to 0$ occurs when the kink is on the top
of a junction.  In this limit the nearest phases can have
small corrections.
We let $\varphi_{j-1}=\alpha$, $\varphi_{j}=\pi-\beta$,
and $\varphi_{j+1}=2\pi-\gamma$.
Again we substitute the corrections and set all the other
phases to 0 or $2\pi$.
Minimizing the energy
with respect to $\alpha$, $\beta$, and $\gamma$ and
linearizing the sine terms yields:
$\alpha={\lambda_{j-1}(\pi-\beta) /(h_{j-1}+ \lambda_{j-2}+\lambda_{j-1})}$,
$\gamma={\lambda_{j}(\pi+\beta) / (h_{j+1}+\lambda_{j}+\lambda_{j+1})}$,
and $\beta$ that can be calculated from
$\lambda_{j-1}(\pi-\beta-\alpha)-\lambda_j(\pi-\gamma+\beta)+h_j\beta=0$.
If we let every $\lambda_j$ be of $O(\lambda)$ and
$O(\lambda)\ll O(h_j)$, then we can expand
the energy as a  series 
\begin{equation}
H=2h_j+ O(\lambda).
\end{equation}
For small $\lambda$, $h_j$ determines the pinning potential height when
the kink is on top of a junction.

The above calculation gives some intuition on the different
ways of designing a ratchet pinning potential. For
instance, alternating critical currents in the array will not produce
a ratchet pinning potential since the potential will still have 
reflection symmetry. In this paper we have
experimentally studied one possible way of breaking this
reflection symmetry by using
alternate critical currents and plaquette areas. However,
another possibility corresponds to having three critical currents
while maintaining equal areas for all the cells. 

\begin{figure}[t]
\epsfxsize=3in
\centering{\mbox{\epsfbox{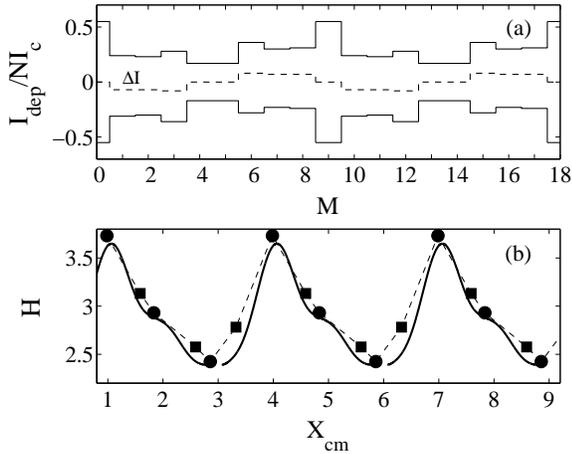}}}
\vspace{0.1in}
\caption[]{(a) Simulated depinning currents for $N=9$ ring
with $\lambda_j=0.25$ and $h_{j-1}=1$, $h_{j}=0.5$, and
$h_{j+1}=0.25$.  Solid lines are the depinning current
as current is increased or decreased while the dashed line is the 
difference of the up and down depinning current.
(b) Numerically calculated pinning
potential. Symbols are analytical calculation
of the energy when a kink is in a plaquette (squares) and on
a junction (circles). The actual kink position is calculated
using {\protect Eq.~(\ref{eq:kinkcm})}.
Dashed line is a guide to the eye.
}
\label{fig:sim_rat_ic}
\end{figure}

To test theses ideas, we have numerically integrated Eq.~\ref{eq:insge} 
for the case of a 9 junctions array.
We let $h_{j-1}=1,h_{j}=0.5,$ and $h_{j+1}=0.25$ (with
$h_{j+3}=h_j$) and use the experimentally realizable value of
$\lambda_j=0.25$ for all $j$.  We set the
kink number $M$ and the initial conditions as described
in the previous section.
We then sweep the applied
current in both the positive and negative direction to
calculate the depinning current.
Fig.~\ref{fig:sim_rat_ic}(a) shows the result of the simulation.

There are three features in the depinning current vs. $M$
graph.  First, the kink is traveling in a ratchet pinning potential.  For
$M=1$, as the current is swept in the positive direction the depinning
current is different than when it is swept in the negative direction.
Second, the depinning current has the expected odd inversion symmetry; that is
$I_{dep+}(M)=I_{dep-}(T-M)$.  Thirdly, the depinning current is
periodic with period $T=9$. All these features were predicted by the
theory developed above.

The observation that the kink is traveling in a ratchet pinning potential can
be directly verified by calculating the pinning potential.  We will use
both the analysis described above and the numerical
method used in \cite{falo99}, which allows us to compute the energy of
the kink as it moves from a maximum to a  minimum.
The position of the kink in the array is calculated with
\begin{equation}
X_{cm}={1\over 2}+{1\over 2\pi}\sum_{j=1}^{N}j(\varphi_{j+1}-\varphi_j).
\label{eq:kinkcm}
\end{equation}
In Fig.~\ref{fig:sim_rat_ic}(b) we have plotted the numerically
calculated pinning potential.
We place the kink
on the energy maximum and perturb it along the unstable
direction and calculate the energy using
Eq.~(\ref{eq:dsgen})  and the kink position using Eq.~(\ref{eq:kinkcm}).
We have also superimposed
the values of the kink pinning potential calculated from the above analysis.  
We have used the linearized results to calculate the phases and
Eq.~(\ref{eq:dsgen}) to calculate the energy.  The circles represent the
energy when the kink is approximately on a junction 
while the squares are the energy when the kink is 
approximately in a plaquette center.
We see that the pinning potential is indeed asymmetric and that
the analysis agrees well with the numerical result.

\section{Summary}
\label{sec:conc}

We have shown that an inhomogeneous parallel
Josephson-junction array provides an ideal experimental system to
study kink motion in different potentials.
In particular, we have
designed a ratchet potential in 
an array with a ring geometry.  One way of designing a
ratchet potential is by varying cell inductances and junction areas.
We have verified experimentally and numerically that a kink, and even
a train of kinks, requires a different amount of force to depin in
positive and negative directions.  One interesting result
for the inhomogeneous rings
is that the periodicity in $M$ of the system will depend
only  on the inductance ratios of consecutive cells.  
As a consequence, it is possible to design
a small ring, e.g. $N=8$, such that one can distinguish
between hundreds of states with different
number of trapped kinks.

We have also shown that a ratchet kink potential can be obtained by
using junctions with 
three different critical currents. In this case, the inductances of all cells
are equal and the array has a
period in $M$ equal to the number of junctions.  

We expect to
investigate a kink in our ratchet potential with an ac bias to show
that there is a rectifying effect: the ac force leads to kink drift in
a preferred direction.  This Brownian rectifier has the added technical
benefit that the dc voltage response is quantized \cite{falo99}.  This
opens up the possibility of designing electronic detectors that can directly measure
the amplitude (instead of just the frequency) of an applied signal very accurately.

The ideas studied in this paper can be extended to the study of vortex
depinning, vortex motion and flux flow in ratchet 2D Josephson-junction arrays. 
We just need to design a 2D array with an appropriate
combination of critical currents and cell areas in the direction
of vortex motion, which is
perpendicular to the current injection direction.

Another way of designing a ratchet effect is by controlling the critical 
current of the individual
junctions of a regular homogeneous array  with the application
of an external magnetic field.  In this way, we
can make a physical realization of a ``flashing ratchet''. 
The mechanics of motion is
well understood \cite{astumian97,hangi}. The pinning potential is removed
periodically. In the interval in which the potential is
off, particles can diffuse freely. After restoration of the pinning
potential, most of the particles localize again in the minimum of the next
lattice site giving a net motion (in the opposite direction of the
``rocking ratchet''). However, as we have seen temperature
(i.e., diffusion) does not play an important role in the motion of the
kink. Nevertheless, one can devise a new mechanism for the kink motion in this
context. After the removal of the pinning potential, kinks delocalize
in an asymmetric way and localize again (when the pinning potential appears) in
the next plaquette. Preliminary numerical simulations \cite{falo2}
confirm this scenario.

The study of inhomogeneous 1D arrays of Josephson junctions can also help to
elucidate pinning mechanism in both 2D Josephson-junction arrays and superconducting
thin films. Also, systems in which critical currents are modulated \cite{fda-suecos} 
can show complex and interesting dynamical behavior. In these systems and
mainly in the presence of ac driving, we expect the appearance of new
collective coherent vortex motion which can give a mode-locking response.
Thus, these ratchet arrays may be used as inspiration
for devices that take advantage
of the properties of directional transport, rectification, and 
quantized response to ac driving.

An interesting application of directional motion of vortices has already been
proposed in \cite{lee99}. An appropriate ratchet potential (via the
modulation of the thickness of the superconductor) is used to eliminate vortices from
the thin film. This ``cleaning'' is also convenient in 1D and 2D 
Josephson-junction arrays in which the
presence of trapped flux breaks the phase coherence of, for instance,
arrays used as radiation sources or complex 
rapid single flux quantum (RSFQ)
circuits.  It appears that our 
ratchet pinning potential could be used to ``clean'' this trapped flux.

In summary, we have shown that inhomogeneous 
parallel arrays of Josephson arrays are ideal model systems for the study of
flux pinning. We have also shown that there are different 
ways to build a ratchet pinning potential,
and have found  an excellent agreement between experiments and theory.

\section*{Acknowledgments}

We thank S.~Watanabe, 
J.E.~Mooij, S.~Cilla, L.M.~Flor\'{\i}a and P.J.~Mart\'{\i}nez 
for insightful discussions.
This work was supported by NSF grant DMR-9610042 and
DGES (PB95-0797 and PB98-1592). JJM thanks the Fulbright Commission and
the MEC (Spain) for financial support.

\section*{Appendix}

In this appendix we calculate the periodicity in the number of kinks, $M$,
of Eq.~(\ref{eq:insge}) for a general inhomogeneous ring array. 
Importantly, this period depends
only on the ratio between consecutive $\lambda's$ and it is independent of
the order of such ratios and the values of the critical currents.

As in the main text, we will use the following transformation for the phases:
\begin{eqnarray}
\psi_j=\varphi_j + 2\pi m_j,
\end{eqnarray}
where $m_j$ is an integer. Eq.~(\ref{eq:transeq}) is the new
equation of motion in the new variables.
The new boundary condition for the transformed variables becomes
$\psi_{j+N}=\psi_j + 2\pi (M+T)$ with
$T=m_{j+N}-m_j$.  The strategy to calculate the period
$T$ will be to find a set of integers that eliminate the
$m_j$ dependence in the right hand side of Eq.~(\ref{eq:transeq}).
We will look for
solutions where $m_j-m_{j-1}+(m_j-m_{j+1})\lambda_j / \lambda_{j-1}=0$.
Clearly, this condition is independent of $h_j$ and only depends on the
ratios $\lambda_j / \lambda_{j-1}$.

First we let $\lambda_j / \lambda_{j-1}=p_j/q_j$ with
$p_j$ and $q_j$ coprime.  Since
only differences of $m_j$ are needed, we let $m_1=0$ without
loss of generality.  Then we
solve for $m_3$ in terms of $m_2$,
\begin{equation}
m_3={p_2+q_2 \over p_2}m_2.
\end{equation}
Similarly, $m_4$ in terms of $m_3$ is
\begin{equation}
m_4={q_2q_3+p_3(p_2+q_2)  \over p_3(p_2+q_2)}m_3.
\end{equation}
After some algebra we find the following recursive formula for
$m_{j+1}/m_{j}$:
\begin{equation}
m_{j+1}/m_j=\xi_{j+1}/p_j \xi_{j},
\end{equation}
with
\begin{equation}
\xi_j=\prod_{k=2}^{j-1}q_k+p_{j-1}\xi_{j-1}.
\end{equation}
Here $\xi_2=1$ and $j=3$ to $N+1$.  
We have now derived that the ratio of
$m_{j+1}/m_j$ is a ratio of integers.  So in
principle, we can find an integer for every $m_j$.

To find a set of integers for $m_j$ we start at the most complex
ratio: $m_{N+1}/m_N$.
We take $m_{N+1}=\xi_{N+1}$ and $m_{N}=p_N\xi_{N}$.
By back substituting, we find 
\begin{equation}
m_j=\xi_{j} \prod_{k=j}^{N}p_k
\end{equation}
for $j=2$ to $N$ and with $m_1=0$.  

It is straight forward to find the
period.
Since we have taken $m_1=0$ the period can be most easily 
expressed as $T=m_{N+1}$,
\begin{equation}
T={\prod_{k=2}^N q_k } + p_N\xi_N.
\label{eq:genper}
\end{equation}
For consistency we also check that the equations
at $j=1$ are satisfied:
$m_0+ m_2 \lambda_1/\lambda_N=0$.  It
is relatively easy to find that $m_0=-\prod_{k=2}^Nq_k$.
The period calculated using $T=m_N-m_0$ also yields
Eq.~(\ref{eq:genper}).  This completes the
existence prove that 
an inhomogeneous parallel array with
consecutive $\lambda's$ that are
rational numbers has a period in $M$.

This procedure, however, will not necessarily yield the minimum
period.  To calculate the minimum period we need to find the 
smallest $m_{N+1}$.  For each ratio of $m_j$, we can make
the numerator and denominator of $\xi_{j+1}/p_j \xi_{j}$ 
relatively prime by dividing by their common multiples.  We start
with the last ratio $m_{N+1}/m_{N}=\xi_{N+1}/p_N \xi_{N}$.  If
we let $y=\gcd(\xi_{N+1},p_N \xi_{N})$ then
$m_{N+1}=\xi_{N+1}/y$ and $m_{N}=p_j \xi_{N}/y$.
However, we also need to be able to consistently change $m_{N}$.
That is, the ratio $m_N/m_{N-1}=p_N\xi_{N}/p_N p_{N-1} \xi_{N-1}$
should still be valid.  This implies that $y$ has
to be a multiple of $m_{N-1}$ as well.  By iterating, we see that
$y$ has to be a multiple of all the $m_j$.  Therefore, let
$x=\gcd(m_{N+1},m_N,\ldots,m_2)$.
The minimum integer period is then
\begin{equation}
T=({\prod_{k=2}^N q_k}+p_N\xi_N)/x.
\end{equation}

As an example, let us consider the regular ring  with
$\lambda_j=\lambda$.  Here
$\xi_{N+1}=N, \gcd(N,N-1, N-2,\ldots, 1)=1$ and $T=N$ as expected
from the homogeneous sine-Gordon equation.  This explains the
observation in Fig.~\ref{fig:sim_rat_ic} that T=9.

As another example, we consider the ring with alternating
areas.  Let $\lambda_j/\lambda_{j-1}=p/q$
for $j$ even and $\lambda_j/\lambda_{j-1}=q/p$ for $j$ odd.  
Then $\xi_3=p+q$, $\xi_4=2pq+q^2$, and
\begin{equation}
\xi_N={N\over 2}p^{N/2-1}q^{{N/ 2}-1}+
({N\over 2}-1)p^{N/2-2}q^{N/2}
\end{equation}
for $N$ even.  Also 
\begin{equation}
\prod_{k=2}^N q_k=p^{N/2-1}q^{N/2}.
\end{equation}
Then, $x=\gcd(m_{N+1},m_N,\ldots,m_2)=p^{N/2-1}q^{N/2-1}$
and
\begin{eqnarray}
T&=&(N/2)p+(N/2-1)q+q \nonumber \\
&=&(p+q)N/2.
\end{eqnarray}
We have recovered the same result derived in the main text.


\begin{references}

\bibitem{braiman95} Y. Braiman, J.~F. Lindner, and W.~L. Ditto, Nature
{\bf 378}, 465 (1995).

\bibitem{sr} P. Jung, Phys.\ Rep. {\bf 234}, 175 (1994);
L. Gammaitoni, P. H\"anggi, P. Jung and F. Marchesoni,
Rev.\ Mod.\ Phys. {\bf 70}, 223 (1998).

\bibitem{astumian97} R.~D. Astumian, Science {\bf 276}, 917 (1997);
F. J\"ulicher, A. Ajdari, J. Prost, Rev.\ Mod.\ Phys.\ {\bf 69},
1269 (1997).

\bibitem{magnasco93} M.~O. Magnasco, Phys.\ Rev.\ Lett. {\bf 71}, 1477
(1993);

\bibitem{hangi} R. Bartussek, P. H\"anggi, and J.G. Kissner
Europhys. Lett., {\bf 28}, 459 (1994); P. H\"anggi and R. Bartussek in
Nonlinear Physics of Complex Systems, Lectures Notes in Physics, Vol
476 (Springer, Berlin) 294-308 (1996).

\bibitem{prost94} J. Rousselet, L. Salome, A. Ajdari and J. Prost,
Nature {\bf 370}, 446 (1994).

\bibitem{lee99} C.S. Lee, B. Janko, I. Der\'enyi and A.L. Barabasi,
Nature {\bf 400}, 337 (1999).

\bibitem{derenyi98} I. Der\'enyi, Ch. Lee and A.L. Barabasi,
Phys.\ Rev.\ Lett. {\bf 80}, 1473 (1998).

\bibitem{FK} See L.~M. Flor\'{\i}a and J.~J. Mazo, Adv.\ Phys. {\bf 45}, 505 
(1996) and references therein.

\bibitem{par} 
A.~V. Ustinov, M. Cirillo and B.~A. Malomed, Phys.\ Rev.\ B {\bf 47},
8357 (1993);
S. Watanabe, H.~S.~J. van der Zant, S.~H. Strogatz and T.~P. Orlando,
Physica D {\bf 97}, 429 (1996).

\bibitem{resnick81} D.~J. Resnick, J.~C. Garland, J.~T. Boyd, S. Shoemaker and
R.~S. Newrock, Phys.\ Rev.\ Lett. {\bf 47}, 1542 (1981).

\bibitem{falo99} F. Falo, P.~J. Mart\'{\i}nez, J.~J. Mazo, and
S. Cilla, Europhys.\ Lett. {\bf 45}, 700 (1999).

\bibitem{fq} To derive the circuit equations we need to apply both
Kirchhoff's current law (KCL) at the nodes and Kirchhoff's voltage law
(KVL) for each loop.  In circuits with Josephson junctions, KVL is
superseded by the more stringent requirement of flux quantization.
Flux quantization gives a constraint on the flux of a loop while KVL
only constrains the derivative of the flux, i.e.  the voltage.
Satisfying flux quantization automatically satisfies KVL.

\bibitem{w1} We have used the Josephson-voltage relation $v =
(\Phi_0/2\pi){\dot \varphi}$ and set the damping ($\Gamma$) to
$\sqrt{\Phi_0/ 2\pi I_c^\star C_\star R_\star^2 }$.  The damping is
usually referred to as the Stewart-McCumber parameter
$\beta_c=\Gamma^{-2}$.

\bibitem{open} In the case of an open array, where open boundary
conditions must be imposed, $f_1 = -2 \pi \lambda_1 A_1 f/A_\star$,
$f_{N}=2\pi \lambda_{N-1} A_{N-1} f/A_\star$ and $f_j=0$ otherwise.

\bibitem{sg} There are several ways of calculating the periodicity in
$M$ of the sine-Gordon equation (regular array).  Probably the most
straight forward is to find a transformation that moves the $M$
dependence from the boundary conditions to the sine term.  In the case
of the regular discrete sine-Gordon equation the new phases are just
$\psi_j=\varphi_j - 2\pi (M / N)j$.
Then $M$ appears only in $\sin(\psi_j + 2\pi {M \over N}j)$ and the
periodicity of the equations in $M$ is just $T=N$.  Though this
approach can also be used in an inhomogeneous rings, it is not trivial to guess
the correct transformation.

\bibitem{dcv} Due to the convex (i.e. inductive) character of the inter-junction coupling
all the junctions show the same dc IV curve.

\bibitem{ambegaokar63} V. Ambegaokar and A. Baratoff, Phys.\ Rev.\
Lett. {\bf 10}, 486 (1963).

\bibitem{fasthenry}
{FastHenry, see http://rle-vlsi.mit.edu}.
See also E. Tr\'{\i}as, Ph.D. thesis, Massachusetts
Institute of Technology, 1999.

\bibitem{baesens}
This correspond to a perturbative expansion from the so-called
anti-integrable limit ($\lambda_j=0$), which can be rigorously justified
(C. Baesens, private communication). For another example see S. Kim, C.
Baesens and R.~S. Mackay, Phys.\ Rev.\ E {\bf 56}, 4955 (1997).

\bibitem{jung96}
See, for example, P. Jung, J. G. Kissner and P. H\"anggi Phys.\ Rev.\ Lett. {\bf 76}, 3436
(1996), for underdamped effects in a ratchet potential.

\bibitem{falo2} Unpublished results.

\bibitem{fda-suecos} F. Dom\'{\i}nguez-Adame, A. S\'{a}nchez and Y.S Kivshar, Phys.
Rev.E {\bf52}, 2183 (1995); E. Lennholm and M. Hornquist, Phys. Rev.E {\bf59},
381 (1999).  

\end{references}
\end{document}